\RequirePackage{fix-cm}
\documentclass[smallextended]{svjour3}       
\smartqed  
\usepackage{graphicx}
\begin{document}

\title{Thermalization of a weakly interacting Bose gas in a disordered trap
}


\author{Che-Hsiu Hsueh \and Makoto Tsubota \and Wen-Chin Wu$^*$
}


\institute{Che-Hsiu Hsueh \at
              Department of Physics, National Taiwan Normal University, Taipei 11677, Taiwan
           \and
           Makoto Tsubota \at
              Department of Physics, Osaka City University, Sugimoto 3-3-138, Sumiyoshi-ku, Osaka 558-8585, Japan and\\
              The OCU Advanced Research Institute for Natural Science and Technology (OCARINA), Osaka, Japan
           \and
           Wen-Chin Wu \at
              Department of Physics, National Taiwan Normal University, Taipei 11677, Taiwan \\
              Tel.: +886-2-77346037\\
              Fax: +886-2-29326408\\
              \email{wu@ntnu.edu.tw}
}

\date{Received: date / Accepted: date}

\maketitle

\begin{abstract}
Previously we numerically showed that thermalization can occur in an oscillating Bose-Einstein condensate
(BEC) with a disordered harmonic trap when healing length $\xi$ of the condensate is shorter than the
correlation length $\sigma_{D}$ of the disorder [see, for example, the experiment reported in
Phys. Rev. A \textbf{82}, 033603 (2010)]. In this work, we investigate the weakly interacting or
Anderson localization regime $\xi>\sigma_{D}$ and show that the oscillating BEC can also
exhibit a relaxation process from nonequilibrium to equilibrium. In such an isolated quantum system,
energy and particle number are conserved and the irreversible evolution towards thermodynamic equilibrium
is induced by the disorder. The thermodynamic equilibrium is evidenced by the maximized entropy
$S\left[n_{k}\right]$ in which the waveaction spectrum $n_{k}$ follows the Rayleigh-Jeans distribution.
Besides, unlike a monotonic irreversible process of thermalization to equilibrium,
the Fermi-Pasta-Ulam-Tsingou recurrence arises in this system,
manifested by the oscillation of the non-equilibrium entropy.

\keywords{Thermalization \and Disorder \and Irreversible process \and Anderson localization}
\end{abstract}

\section{Introduction}
\label{intro}
One of the longstanding questions in statistical mechanics is that how an isolated system can relax
from nonequilibrium to equilibrium, {\em i.e.}, thermalization
\cite{nature04693,nature06838,RevModPhys.83.863,BORGONOVI20161}.
This question also connects to how occurs the second law of thermodynamics
and what orientates the direction of time. In one our recent paper \cite{RefJ_Hsueh2018PRL},
we numerically simulate an oscillating Bose-Einstein condensate (BEC) in a
disordered trap \cite{RefJ_Hulet2010PRA} and the results are in good agreement with the experiment.
It showed that when the healing length
$\xi$ of the condensate is shorter than the correlation length $\sigma_{D}$ of the disorder,
the system can eventually approach a thermodynamic equilibrium that
is accompanied by an {\em algebraic} localization.
In this paper, we show that an oscillating BEC in the weakly interacting or Anderson localization regime,
$\xi>\sigma_{D}$, can also exhibit a relaxation process from nonequilibrium to equilibrium.

In the literature, most of works concerning the thermalization in a nonlinear system
focused on the condensation process of nonlinear waves with a given
incoherent initial condition \cite{RefJ_Picozzi2005PRL,RefJ_Picozzi2007OE,RefJ_Picozzi2012NP}.
In these systems, the interaction between particles plays the central role towards the thermalization.
In this work, in contrast, we focus on a condensate with a {\em moving} coherent initial state where
the thermalization arises due to the random disorder. It is worth noting that
for the present system the disorder does not cause any energy dissipation.
It acts as  a medium which results in the exchange of partial kinetic energy
with partial potential energy and the subsequent equilibrium.
As a consequence, random disorder is realized as another route towards the thermalization.

\section{Theoretical Approach}
\label{model}
We consider a 1D Bose gas with a repulsive contact interaction that is confined in a harmonic potential $V_{h}\left(z\right)=m\omega^{2}z^{2}/2$. In the dilute and ultracold condition,
the condensate wave function $\psi\left(z,t\right)$, normalized to one,
$\int\left|\psi\right|^{2}dz=1$, is governed by the Gross-Pitaveskii (GP) equation in
the presence of a real spatially random disordered potential $V_{D}\left(z\right)$,
\begin{equation}\label{disorderGP}
  i\hbar\partial_{t}\psi\left(z,t\right)=\left[-\frac{\hbar^{2}\partial_{z}^{2}}{2m}
  +V_{h}\left(z\right)+V_{D}\left(z\right) \\
  +g\left|\psi\left(z,t\right)\right|^{2}\right]\psi\left(z,t\right).
\end{equation}
The healing length at the center of the condensate is defined as $\xi=\hbar/\sqrt{2m\mu}$,
where $\mu=\left(3g/2^{5}\right)^{2/3}$ is the chemical potential
with $g$ the coupling constant of contact interaction. The disorder correlation length $\sigma_{D}$
is defined by fitting the autocorrelation function
$\langle V_{D}(z)V_{D}(z+\Delta z)\rangle=V_{0}^{2}\exp\left(-2\Delta z^{2}/\sigma_{D}^{2}\right)$
with $V_{0}$ the strength of $V_{D}\left(z\right)$. In the experiment \cite{RefJ_Hulet2010PRA},
the condensate is released at a position off the center of the harmonic trap that results in
the subsequent oscillations. For numerical convenience,
we take the alternative scheme such that the condensate is released at the trap center
but with an initial velocity $v_{0}$. To obtain an initial wave function with a velocity
$v_{0}$, we apply the Galilean transformation: $\psi=\varphi\exp\left(imv_{0}z\right)$ and in
the absence of disorder, the corresponding GP equation for the residual wave function $\varphi$ is
\begin{equation}\label{GP}
  i\hbar\partial_{t}\varphi\left(z,t\right)=\left[\frac{1}{2m}\left(\frac{\hbar}{i}\partial_{z}
  -mv_{0}\right)^{2}+V_{h}\left(z\right) \\
  +g\left|\varphi\left(z,t\right)\right|^{2}\right]\varphi\left(z,t\right).
\end{equation}
Long-term imaginary-time evolution of Eq. (\ref{GP}) gives $\varphi$ which in turn gives the
initial wave function $\psi$. The experimental parameters are $\omega=2\pi\times5.5$ Hz,
$\mu=h\times1.1$ kHz, $V_{0}/h=280$ Hz, and oscillation amplitude
$A=0.6$ mm which yields an initial peak velocity
$v_{0}=20$ mm/s. Throughout the calculations, we use $l_{h}=\sqrt{\hbar/m\omega}$
and $\tau_{h}=\omega^{-1}$ as the units of length and time respectively.
The experiment \cite{RefJ_Hulet2010PRA} was performed using $\sigma_{D}=0.25 l_{h}$
that falls into the regime $\xi < \sigma_{D}$. Here we instead choose
$\sigma_{D}=0.01l_{h}$ that corresponds to the regime $\xi>\sigma_{D}$
of Anderson localization \cite{RefJ_Sanchez-PalenciaPRL}.

Thermalization is intimately related to the flux, energy, or particle waveaction
between the microstates.
Owing to the random disorder, the wave function $\psi$ is seen to exhibit an irreversible
evolution towards thermal equilibrium. Thermalization phenomenon can be studied in terms of
effective diffusion in the momentum space.
Such salient properties of energy or waveaction flux is well described in the context of
wave-turbulence (WT) theory. In this work, we shall apply the
WT theory to study the equilibrium properties of the system. In this regard,
it is useful to express the condensate wave function in
form of Madelung transformation, $\psi(z,t)=\sqrt{\rho(z,t)}\exp[i\phi(z,t)]$
with $\rho$ and $\phi$ the density and phase, respectively.
As a result, total energy of the system can be expressed as the sum of five terms,
$E_{tot}(t)=E_{hyd}(t)+E_{qum}(t)+
E_{pot}(t)+E_{dis}(t)+E_{int}(t)$, where $E_{hyd}=(\hbar^{2}/2m)\int\rho|\nabla\phi|^{2}dz$
is the hydrodynamic kinetic energy, $E_{qum}=(\hbar^{2}/2m)\int|\nabla \sqrt{\rho}|^{2}dz$ is
the quantum pressure energy, $E_{pot}=\int\rho V_{h}dz$ is the trapping energy,
$E_{dis}=\int\rho V_{D}dz$ is the disorder-potential energy,
and $E_{int}=(g/2)\int\rho^{2}dz$ is the interaction energy.
To characterize the energy flux, we employ the sum rule in wave number $k$ space:
$E_{hyd}\left(t\right)=\int_{0}^{k_{c}}\mathcal{E}_{hyd}\left(k,t\right)dk$, where
$k_{c}$ is an ultraviolet cutoff and the hydrodynamic kinetic energy spectrum
\begin{equation}\label{kinetic}
  \mathcal{E}_{hyd}\left(k,t\right)=\frac{\hbar^{2}}{m}\left|\int\exp\left(-ikz\right)
  \sqrt{\rho\left(z,t\right)}\partial_{z}\phi\left(z,t\right)dz\right|^{2}.
\end{equation}
We will elaborate the cutoff $k_{c}$ later.

\begin{figure}
\begin{center}
\includegraphics[width=0.75\textwidth]{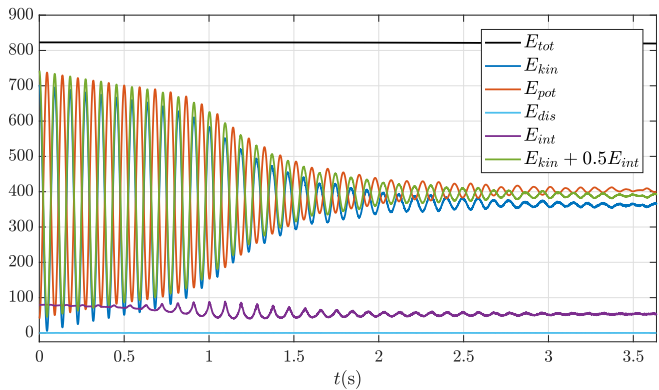}
\end{center}
\caption{Time evolution of four energies $E_{kin}$, $E_{pot}$, $E_{dis}$, and $E_{int}$ discussed
in text. Total energy is conserved and $E_{dis}\simeq0$ for the entire process.
$E_{kin}$, $E_{pot}$, and $E_{int}$ all come to a constant at $t\rightarrow\infty$ which
evidences that the system is approaching the equilibrium.
Comparison of the red and green curves reveals that the condition of the virial theorem,
$E_{pot}\simeq E_{kin}+E_{int}/2$, is satisfied at equilibrium.
$E_{kin}(t\rightarrow \infty)\equiv k_B T/2\approx 364\hbar\omega$ defines the equilibrium
temperature $T\approx0.53$ nK.}
\label{fig:1}       
\end{figure}

\section{Results and Discussions}
\label{results}
\begin{figure}
\begin{center}
\includegraphics[width=0.75\textwidth]{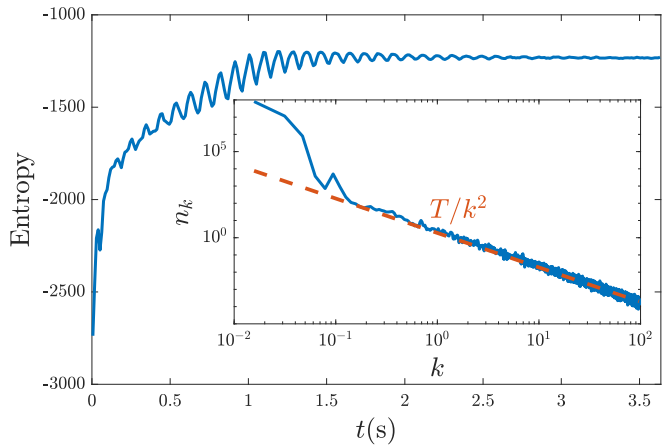}
\end{center}
\caption{Temporal entropy (\ref{entropy}) of the oscillating condensate. The inset shows
the corresponding waveaction (\ref{waveaction}) at t = 3.6s,
that is well fitted by the Rayleigh-Jeans spectrum (\ref{RJ}).
The temperature $T$ of the fitting (red) curve agrees
with the equilibrium $E_{kin}\approx364\hbar\omega$ in Fig.~\ref{fig:1}.}
\label{fig:2}       
\end{figure}

Virial theorem describes the relationship between various energies of a system
when it approaches the equilibrium. For the current system described by the GP
equation (\ref{disorderGP}), according to the virial theorem
the following condition should be satisfied at equilibrium \cite{RefB_Pitaevskii2016},
\begin{equation}\label{virial}
  2E_{kin}-2E_{pot}+dE_{int}=0.
\end{equation}
Here $E_{kin}$ is the quantum kinetic energy consisting of both hydrodynamic kinetic energy and
quantum pressure energy, $E_{kin}=E_{hyd}+E_{qum}$,
and $d$ denotes the dimension. Fig.~\ref{fig:1} shows the time evolution
of the four energies, $E_{kin}$, $E_{pot}$, $E_{dis}$, and $E_{int}$.
As random disorder potential is rapidly varying in space, $E_{dis}\simeq0$ (cyan line)
for the entire process. Both the trapping and disorder potentials
are real and time-independent, thus one expects that there is no energy loss.
Conservation of total energy is indeed shown in Fig.~\ref{fig:1} (black line).
Of most interest, during the process partial kinetic energy is in exchange with
partial potential energy and all $E_{kin}$, $E_{pot}$, and $E_{int}$
come to a constant when $t\rightarrow\infty$ that signals
the equilibrium. Moreover, it is confirmed that
the condition (\ref{virial}) of virial theorem, $E_{pot}\simeq E_{kin}+E_{int}/2$,
is satisfied for the current system with $d=1$
(see the comparison between the red and green lines in Fig.~\ref{fig:1}).
One can define the equilibrium temperature $T$ from the kinetic energy at equilibrium,
$E_{kin}(t\rightarrow \infty)\equiv k_B T/2\simeq 364\hbar\omega$. It gives
$T\simeq 0.53$ nK which will be used to fit the waveaction spectrum in Fig.~\ref{fig:2}.

During the thermalization process, there are energy or waveaction flux between
the microstates. The waveaction spectrum can be given as \cite{RefB_Nazarenko2011}
\begin{equation}\label{waveaction}
  n_{k}=\frac{m}{\hbar^{2}k^{2}}\mathcal{E}_{hyd}\left(k,t\right),
\end{equation}
where $\mathcal{E}_{hyd}$ is the hydrodynamic kinetic energy spectrum defined in (\ref{kinetic}).
In terms of $n_{k}$, one can further define an entropy
\cite{RefJ_Picozzi2005PRL,RefJ_Picozzi2007OE,RefJ_Picozzi2012NP}
\begin{equation}\label{entropy}
  S\left(t\right)=\int\ln n_{k}dk.
\end{equation}
For WT, the simplest steady state is the one corresponding to no flux of energy or waveaction
between the microstates -- so called the Rayleigh-Jeans (RJ) spectra.
Thus RJ spectrum corresponds to a state of energy equipartition in the $k$-space, or
a non-dissipative state of detailed balancing for
the local energy transfer.
For the current system, the corresponding RJ spectrum appears to be
\begin{equation}\label{RJ}
  n^{\rm RJ}=\frac{T}{k^{2}+\mu},
\end{equation}
where $T$ denotes the equilibrium kinetic energy or temperature
(see Fig.~\ref{fig:1}) and $\mu$ is the chemical potential.

In a real system, the RJ spectra can only be realized with a truncated $k_c$ in the $k$-space,
however. That is, only a large but finite number of modes are involved in the RJ spectra.
Such phenomenon is known as \emph{ultraviolet catastrophe}.
How a ultraviolet cutoff $k_{c}$ can make the RJ spectrum a legitimate solution
is a real issue. In our previous simulation \cite{RefJ_Hsueh2018PRL} on
the dipole oscillation experiment with $\xi <\sigma_D$ \cite{RefJ_Hulet2010PRA},
the cutoff $k_c$ which gives the best fit is found to be slightly smaller than the
wave number corresponding to the healing length, $k_{\xi}\equiv 2\pi/\xi$.
It indicates that the optimal cutoff $k_c$ actually mimics the
shortest length scale or the largest $k$ scale in association with
the healing length of the system. In the current case with $\xi >\sigma_D$,
we shall use the similar cutoff $k_c$.

Fig.~\ref{fig:2} shows the evolution of the entropy (\ref{entropy}).
The inset shows the eventual waveaction spectrum (\ref{waveaction}) that is well fitted
by the RJ spectrum (\ref{RJ}).
It unambiguously indicates that thermalization is developing in the system.
In earlier process, one also sees clear Fermi-Pasta-Ulam-Tsingou (FPUT)
recurrence effect \cite{FORD1992271,Zaslavsky2005} that is consistent with the oscillation of the system.

Finally we derive the relevant equation of motion for the irreversible evolution
of the system. The GP equation (\ref{disorderGP}) can be transformed to in the momentum space:
\begin{equation}\label{disorderGP_k}
  i\hbar\partial_{t}\widetilde{\psi}_{k}=\widehat{H}_{0}\widetilde{\psi}_{k}+
  \frac{g}{\left(2\pi\right)^{2}}\int\widetilde{\psi}_{k_{3}}^{\ast}\widetilde{\psi}_{k_{2}}
  \widetilde{\psi}_{k_{1}}\delta_{12}^{3k}dk_{1}dk_{2}dk_{3}\\
  +\frac{1}{2\pi}\int\widetilde{V}_{D}\left(k-k_{1}\right)\widetilde{\psi}_{k_{1}}dk_{1},
\end{equation}
where $\widehat{H}_{0}=\hbar^{2}k^{2}/2m-m\omega^{2}\partial_{k}^{2}/2$, $\delta_{12}^{3k}
=\delta\left(k_{1}+k_{2}-k_{3}-k\right)$, and
$\widetilde{\psi}_{k}$ and $\widetilde{V}_{D}$ are the Fourier transformations
of the wave function $\psi(z)$ and the disorder potential $V_{D}(z)$, respectively. Writing
$\widetilde{\psi}_{k}\equiv\sqrt{D_{kk}}\exp\left(i\phi_{k}\right)$
or $D_{kk}\equiv\widetilde{\psi}_{k}\widetilde{\psi}_{k}^{\ast}$,
Eq.~(\ref{disorderGP_k}) can be reduced
to a Boltzmann like equation and a coupled hydrodynamic equation:
\begin{equation}\label{BTE}
  \partial_{t}D_{kk}+\frac{\partial_{k}}{\hbar}\left(FD_{kk}\right)=
  \Im\left\{\mathcal{C}\left[D\right]\right\}+\Im\left\{\mathcal{S}\left[D\right]\right\}
\end{equation}
\begin{equation}\label{hyd_k}
  \partial_{t}\phi_{k}=-\frac{\hbar k^{2}}{2m}-\frac{m\omega^{2}}{2\hbar}\left[\left(\partial_{k}\phi_{k}\right)^{2}
  -\frac{\partial_{k}^{2}\sqrt{D_{kk}}}{\sqrt{D_{kk}}}\right]
  -\frac{\Re\left\{\mathcal{C}\left[D\right]\right\}}{D_{kk}}
  -\frac{\Re\left\{\mathcal{S}\left[D\right]\right\}}{D_{kk}}.
\end{equation}
Here $F=m\omega^{2}\left(\partial_{k}\phi_{k}\right)$ corresponds to the
force due to the harmonic trap, $\Im$ and $\Re$ correspond to the imaginary and real parts,
and $\mathcal{C}\left[D\right]$ and $\mathcal{S}\left[D\right]$
correspond to the terms associated with the collision between particles and
the scattering by disorder, respectively. Explicitly
\begin{equation}\label{coll}
  \mathcal{C}\left[D\right]=\frac{g}{2\pi^{2}\hbar}\int D_{k_{2}k_{3}}D_{k_{1}k}\delta_{12}^{3k}dk_{1}dk_{2}dk_{3}
\end{equation}
\begin{equation}\label{scat}
  \mathcal{S}\left[D\right]=\frac{1}{\pi\hbar}\int\widetilde{V}_{D}\left(k-k_{1}\right)D_{k_{1}k}dk_{1},
\end{equation}
where $D_{k_{1}k}\equiv\widetilde{\psi}_{k_{1}}\widetilde{\psi}_{k}^{\ast}$.

In the case without the trap, $F=0$, the Boltzmann like equation (\ref{BTE}) reduces
to a master equation in which the disorder potential
$\widetilde{V}_{D}$ plays the central role of the transition matrix.
In this case, the simplest stationary solution of Eq.~(\ref{BTE}) occurs when all the off-diagonal terms
vanish, $D_{k_{1}k}=0$ for all $k_1\neq k$.
This corresponds to the detailed balancing which gives rise to the RJ spectrum.
In the present case of a smooth trapping potential, $F\simeq 0$
seems to be a good approximation and the above simplest stationary solution for $F=0$ holds.

\section{Conclusions}
\label{conclusion}

We show that an oscillating condensate in a disordered trap is an excellent system
to exhibit a relaxation process from nonequilibrium to equilibrium.
Here we focus on the case when the healing length of the condensate
is exceeding the correlation length of disorder -- so-called the Anderson localization regime.
Due to the random disorder, the oscillating condensate is eventually stopped that signals the
equilibrium.
We have confirmed that when the system comes to the equilibrium, the entropy is maximized and
the waveaction spectrum follows the Rayleigh-Jeans dispersion.
A Boltzmann like equation has been derived to
explain the role of the disorder.

\section*{Acknowledgements}
\label{acknowledgement}

Financial supports from MOST, Taiwan (grant No. MOST 105-2112-M-003-005),
JSPS KAKENHI (grant No. 17K05548) and MEXT KAKENHI/Fluctuation and Structure
(grant No.16H00807), and NCTS of Taiwan are acknowledged.

\end{document}